\documentclass[%
 reprint,
 amsmath,amssymb,
 aps,
]{revtex4-1}

\usepackage{multirow}
\usepackage{mathrsfs,amsmath}
\usepackage{hyperref}
\usepackage{verbatim}
\usepackage{amsmath}
\usepackage{amsfonts}
\usepackage{amssymb}
\usepackage{amsthm}
\usepackage{bm}
\usepackage{xparse}
\usepackage{graphicx}
\usepackage{xcolor}
\usepackage{textcase}
\usepackage{url}
\usepackage{lipsum}
\usepackage{appendix}
\usepackage{epstopdf}
\usepackage{footnote}
\usepackage{color}
\usepackage[Symbol]{upgreek}

\begin{document}

\preprint{APS/123-QED}

\title{Truncated triangular diffraction lattices and orbital-angular-momentum detection of vortex SU(2) geometric modes}

\author{Yijie Shen}
\author{Xing Fu}%
\author{Mali Gong}%
\email{shenyj15@mails.tsinghua.edu.cn}
\affiliation{%
	State Key Laboratory of Precision Measurement Technology and Instruments, Center for Photonic and Electronic, Department of Precision Instruments, Tsinghua University, Beijing, 100084, China}%

\date{\today}

\begin{abstract}
We for the first time report the truncated diffraction with a triangular aperture of the SU(2) geometric modes and propose a method to detect the complicated orbital angular momentum (OAM) of an SU(2) wave-packet, to the best of our knowledge. As a special vortex beam, a nonplanar SU(2) mode carrying special intensity and OAM distributions brings exotic patterns in truncated diffraction lattice. A meshy structure is unveiled therein by adjusting the illuminated aperture in vicinity of the partial OAM regions, which can be elaborately used to evaluate the partial topological charge and OAM of an SU(2) wave-packet by counting the dark holes in the mesh. Moreover, through controlling the size and position of the aperture at the center region, the truncated triangular lattice can be close to the classical spot-array lattice for measuring the center OAM. These effects being fully validated by theoretical simulations greatly extend the versatility of topological structures detection of special beams.
\end{abstract}

\maketitle

\section{Introduction}
Optical vortices carrying orbital angular momentum (OAM) have recently hatched widespread applications such as optical tweezers \cite{1,2}, optical communication \cite{3,4}, nonlinear optics \cite{5,6}, and quantum entanglement \cite{7,8}. It is a traditional method to measure OAM by counting the strips in the interferograms via interferometer experiments. However, this method is generally cumbersome, as well as invalid for some high-power or ultrashort pulsed or special structured vortex beams. There is an everlasting challenge for conveniently and effectively detecting the OAM information including phase-singularities and topological charges of general vortex beams. Since Hickmann et al. \cite{9}, in 2010, unveiled the truncated optical lattice in triangular-aperture diffraction of OAM beams, it has been used as a more effective method for OAM detection. This method is free of complicated interferometer devices and has been successfully used in femtosecond vortices \cite{10}, non-integer charge vortices \cite{11}, vectorial vortices \cite{12}, and elliptical vortex beams \cite{13}, which largely promoted the related technologies and applications. Therefore, it is very significant to fulfill the technical breakthrough by more characteristic structured lights for extending more novel applications.

As a new kind of special vortex beams, nonplanar SU(2) geometric mode \cite{14} has recently arouse great interest due to its peculiar properties like periodic ray trajectory \cite{15,16}, drastic increase of power \cite{16,17}, multiple singularities \cite{18}, and fractional OAM \cite{19}. When a resonator cavity fulfills the reentrant condition of SU(2)-Lie coupled harmonic oscillation, the laser modes undergo frequency degeneracy and the photon wave-packet performs as SU(2) quantum coherent states \cite{20,21}. The exotic modes in the frequency degenerate cavity under off-axis pumping are always called geometric modes \cite{14,15,16,17,18,19}. The nonplanar geometric modes can be obtained by astigmatic transformation of planar geometric modes \cite{14,18} and can also be directly emitted from frequency degenerate cavity \cite{19,22}. The SU(2) geometric modes has been used to explore novel high-pulse-energy vortices \cite{14,19} and structured polarized beams \cite{22,23}. However, it has never been settled on how to experimentally measure the actual OAM of an SU(2) wave-packet. Although the interferometer experiment of SU(2) vortices was illustrated for verifying the existence of OAM and topological charge\cite{18}, yet the interferograms cannot be used to count the actual value of topological charges as the case of a classical doughnut-shape beam. The key issue leading to this difficulty is that the SU(2) vortices have complicated multi-singularity and fractional OAM structure. The exotic singularities distribution gives rise to not only the center dark region but also several partial dark regions, which forbid the emergence of overall strips in conventional interference technique\cite{18,19}. The topological charges exist not only in the center dark region but also in the partial dark regions. To date, it was still a great challenge to quantitatively identify the topological charges in a vortex SU(2) wave-packet. Therefore, it will be a great significance if a quantitatively detecting method of topological phase of an SU(2) vortex can be verified applying truncated optical lattices.

In this work, we for the first time, to the best of our knowledge, demonstrate the truncated diffraction of SU(2) vortices and propose the OAM detecting method valid for both the center and the partial topological charges measurment. By observing the Fraunhofer diffraction of SU(2) vortices through triangular apertures, we found an exotic meshy structure in the truncated lattices pattern, which is correlated with the topological intensity-phase structure of SU(2) wave-packet. When we adjust the size, position, and orientation of the aperture to cover the partial OAM region, the meshy structure can be transformed into a linear dark-holes mesh and the number of the holes is correlated with the partial topological charges. Moreover, with a proper adjustment of the aperture at the center region, the truncated triangular lattice can be close to the classical spot-array lattice for measuring the center OAM. Moreover, we present adequate theoretical simulations for these effects showing good agreement with the experimental results.
\begin{widetext}
	\section{Theories}
	\subsection{Basic equations}
	In this section, we give the basic diffraction modeling of an incident optical vortex on an aperture located at $z=0$ plane, $z$ being the propagation axis. As is known, the basis to constitute the optical vortices is the Hilbert factor $\exp \left( {\rm i}\ell \theta  \right)$. Supposing a unitary vortex field is in front of the aperture, the slowly varying electric field envelope is written as:
	\begin{equation}
	{{E}_{\ell }}\left( x,y,0 \right)={{\left[ x+{\rm i}\cdot {\rm sgn}\left( \ell  \right)y \right]}^{\left| \ell  \right|}}\equiv {{E}_{\ell }}\left( r,\theta ,0 \right)={{r}^{\left| \ell  \right|}}\exp \left( {\rm i}\ell \theta  \right),
	\label{e1}
	\end{equation}
	here we give the expression both in Cartesian and cylindrical polar coordinates necessary for further discussions. Then, supposing an amplitude modulation function $t\left( x,y \right)$ of the aperture, the Fraunhofer diffraction field at Fourier plane $z=f$ is given by:
	\begin{align}
	\nonumber
	{{E}_{\ell }}\left( \xi ,\eta ,f \right)&=\int_{-\infty }^{\infty }{\int_{-\infty }^{\infty }{t\left( x,y \right){{\left[ x+{\rm i}\cdot {\rm sgn}\left( \ell  \right)y \right]}^{\left| \ell  \right|}}{\rm exp}\left[ -\frac{{2}\pi {\rm i}}{\lambda f}\left( \xi x+\eta y \right) \right]{\rm d}x{\rm d}y}} \\ \nonumber
	&=\frac{1}{2\pi }\mathscr{F}\left[ t\left( x,y \right) \right]*\mathscr{F}\left\{ {{\left[ x+{\rm i}\cdot {\rm sgn}\left( \ell  \right)y \right]}^{\left| \ell  \right|}} \right\} \\ \nonumber
	& =\frac{1}{2\pi }T\left( {{\omega }_{x}},{{\omega }_{y}} \right)*\mathscr{F}\left\{ \sum\limits_{k=0}^{\left| \ell  \right|}{\left( \begin{matrix}
		\left| \ell  \right|  \\
		k  \\
		\end{matrix} \right)}{{x}^{\left| \ell  \right|-k}}{{y}^{k}}{{\left[ {\rm i}\cdot {\rm sgn}\left( \ell  \right) \right]}^{k}} \right\} \\ \nonumber
	& =\frac{1}{2\pi }T\left( {{\omega }_{x}},{{\omega }_{y}} \right)*\sum\limits_{k=0}^{\left| \ell  \right|}{\left( \begin{matrix}
		\left| \ell  \right|  \\
		k  \\
		\end{matrix} \right)}{{\left[ {\rm i}\cdot {\rm sgn}\left( \ell  \right) \right]}^{k}}\left[\frac{1}{2\pi }\mathscr{F}\left( {{x}^{\left| \ell  \right|-k}} \right)*\mathscr{F}\left( {{y}^{k}} \right)\right]\\ \nonumber
	& =\frac{1}{4{{\pi }^{2}}}T\left( {{\omega }_{x}},{{\omega }_{y}} \right)* \sum\limits_{k=0}^{\left| \ell  \right|}{\left( \begin{matrix}
		\left| \ell  \right|  \\
		k  \\
		\end{matrix} \right)}{{\left[{\rm i}\cdot{\rm sgn}\left( \ell  \right) \right]}^{k}}\left[ 2\pi {{{\rm i}}^{\left| \ell  \right|-k}}{{\updelta}^{\left( \left| \ell  \right|-k \right)}}\left( {{\omega }_{x}} \right) \right]*\left[ 2\pi {{{\rm i}}^{k}}{{\updelta}^{\left( k \right)}}\left( {{\omega }_{y}} \right) \right]  \\ \nonumber
	& ={{{\rm i}}^{\left| \ell  \right|}}\sum\limits_{k=0}^{\left| \ell  \right|}{\left( \begin{matrix}
		\left| \ell  \right|  \\
		k  \\
		\end{matrix} \right)}{{\left[ {\rm i}\cdot {\rm sgn}\left( \ell  \right) \right]}^{k}}T\left( {{\omega }_{x}},{{\omega }_{y}} \right)*{{\updelta}^{\left( \left| \ell  \right|-k \right)}}\left( {{\omega }_{x}} \right)*{{\updelta}^{\left( k \right)}}\left( {{\omega }_{y}} \right) \\
	& ={{\left( -{\rm i} \right)}^{\left| \ell  \right|}}\sum\limits_{k=0}^{\left| \ell  \right|}{\left( \begin{matrix}
		\left| \ell  \right|  \\
		k  \\
		\end{matrix} \right)}{{\left[ {\rm i}\cdot {\rm sgn}\left( \ell  \right) \right]}^{k}}\frac{{{\partial }^{\left| \ell  \right|}}}{\partial \omega _{x}^{\left| \ell  \right|-k}\partial \omega _{y}^{k}}T\left( {{\omega }_{x}},{{\omega }_{y}} \right),
	\label{e2}
	\end{align}
	where $\lambda$ is light wavelength, $\left( {{\omega }_{x}},{{\omega }_{y}} \right)=\left( \frac{2\pi \xi }{\lambda f},\frac{2\pi \eta }{\lambda f} \right)$ is the  angular frequency, $T\left( {{\omega }_{x}},{{\omega }_{y}} \right)$ is the Fourier transform of aperture function, and $\left( \begin{matrix}
	m  \\
	n  \\
	\end{matrix} \right)$ represents the binomial coefficient. For the specific case that $t\left( x,y \right)$ describes an equilateral triangular aperture with unity transmission inside the aperture and zero outside, the pattern of  ${{\left| {{E}_{\ell }}\left( \xi ,\eta  \right) \right|}^{2}}$ produces a characteristic truncated triangular lattice correlated with the topological charge $\ell$. This kind of lattice has been used as a powerful application for directly detecting the OAM of light \cite{9,10,11,12}.
	
	\subsection{Diffraction of vortex SU(2) wave-packet}
	To fulfill the reentrant condition of coupled harmonic oscillators in SU(2) Lie algebra, the cavity configuration requires a frequency-degenerate state of $\Omega =\Delta {{f}_{T}}/\Delta {{f}_{L}}=P/Q\in \mathbb{Q}$, where $P$ and $Q$ are coprime integers, and $\Delta {{f}_{L}}$ ($\Delta {{f}_{T}}$) is the longitudinal (transverse) mode spacing \cite{20,21}. In frequency-degenerate cavity under off-axis pumping along $x$-axis, it has been proved that the emission mode is performed as SU(2) coherent state \cite{14}:
	\begin{equation}
	\Psi _{{{n}_{0}}}^{M}\left( x,y,z;{{\phi }_{0}}\left| \Omega  \right. \right)=\frac{1}{{{2}^{{M}/{2}\;}}}\sum\limits_{K=0}^{M}{\sqrt{\frac{M!}{K!\left( M-K \right)!}}}\cdot {{\text{e}}^{\text{i}K{{\phi }_{0}}}}\cdot \psi _{{{n}_{0}}+Q\cdot K,0,{{s}_{0}}-P\cdot K}^{\left( \text{HG} \right)}\left( x,y,z \right),
	\label{e3}
	\end{equation}
	where $M\text{+1}$ is the number of Hermite--Gaussian (HG) modes in frequency-degenerate family of $\psi _{{{n}_{0}}+Q\cdot K,0,{{s}_{0}}-P\cdot K}^{\left( \text{HG} \right)}\left( x,y,z \right)$. Therein, ${{n}_{0}}$ and ${{s}_{0}}$ represent the minimum transverse and maximum longitudinal orders in the degenerate family, respectively; the phase factor ${{\phi }_{0}}$ links to the ray trajectory of classical geometric periodic orbits \cite{15}. When $M=0$ in Eq.~(\ref{e3}), the SU(2) wave-packet is reduced to a single HG mode, revealing the common high-order mode in non-degenerate states. The wave function of HG mode is represented as:
	\begin{align}
	\nonumber\psi _{n,m,s}^{\left( \text{HG} \right)}\left( x,y,z \right)&=\sqrt{\frac{2}{\pi m!n!}}\cdot \frac{{{2}^{-{\left( m+n \right)}/{2}\;}}}{w\left( z \right)}{{H}_{n}}\left[ \frac{\sqrt{2}x}{w\left( z \right)} \right]{{H}_{m}}\left[ \frac{\sqrt{2}x}{w\left( z \right)} \right]\exp \left[ -\frac{{{x}^{2}}+{{y}^{2}}}{{{w}^{2}}\left( z \right)} \right] \\ 
	& \text{ }\times \exp \left[ \text{i}{{k}_{n,m,s}}z+\text{i}{{k}_{n,m,s}}\frac{{{x}^{2}}+{{y}^{2}}}{2R\left( z \right)}-\text{i}\left( m+n+1 \right)\vartheta \left( z \right) \right],
	\label{e4}  
	\end{align}
	where ${{k}_{n,m,s}}=2\pi {{f}_{n,m,s}}/c$ with ${{f}_{n,m,s}}$ being the eigenmode frequency and $c$ being the speed of light in vacuum. In the above equations, ${{H}_{n}}(\cdot )$ represents the Hermite polynomials of $n$-th order, $R\left( z \right)={\left( z_{R}^{2}+{{z}^{2}} \right)}/{z}\;$ the radius of curvature where ${{z}_{R}}$ is the Rayleigh range, $\vartheta (z)={{\tan }^{-1}}(z/{{z}_{R}})$ the Gouy phase, $w(z)={{w}_{0}}\sqrt{1+(z/z_R)^2}$ the waist parameter, in which ${{w}_{0}}=\sqrt{\lambda {{z}_{R}}/\pi }$ represents the fundamental mode radius at the waist. The eigenmode frequency of resonator is given by ${{f}_{n,m,s}}=s\Delta {{f}_{L}}+(n+m+1)\Delta {{f}_{T}}$. In a plano-concave cavity with the length of $L$ and the radius of curvature of $R$, the longitudinal mode spacing is $\Delta {{f}_{L}}=c/(2L)$ and the transverse mode spacing is given as $\Delta {{f}_{T}}=\Delta {{f}_{L}}\vartheta (L)/\pi $, therefore, the mode-spacing ratio is $\Omega =(1/\pi ){{\cos }^{-1}}(\sqrt{1-L/R})$. The shape of laser mode in state of $\Omega =P/Q$ has a preference to be localized on the periodic ray trajectories with the period of $Q$. With an external $\pi/2$-astigmatic mode convertor (AMC) \cite{24}, HG modes can be transformed into Laguerre--Gaussian (LG) modes with the generation of OAM, i.e. the transverse mode HG$_{m,n}$ is transformed into LG$_{p,\ell}$ with relationships $p=\min \left( m,n \right)$ and $\ell =\pm \left( m-n \right)$ where the sign of $\pm$ depends on the inclined angle (45$^\circ$ or 135$^\circ$) of HG mode. Consequently, via AMC, the planar geometric mode as described by Eq.~(\ref{e3}) can be transformed into the nonplanar vortex geometric mode:
	\begin{equation}
	\Phi _{{{n}_{0}}}^{M}\left( \rho ,\theta ,z;{{\phi }_{0}}\left| \Omega  \right. \right)=\frac{1}{{{2}^{{M}/{2}\;}}}\sum\limits_{K=0}^{M}{\sqrt{\frac{M!}{K!\left( M-K \right)!}}}\cdot {{\text{e}}^{\text{i}K{{\phi }_{0}}}}\cdot \varphi _{0,\pm \left( {{n}_{0}}+Q\cdot K \right),{{s}_{0}}-P\cdot K}^{\left( \text{LG} \right)}\left( \rho ,\theta ,z \right),
	\label{e5}
	\end{equation}
	where the wave function of LG mode is represented by:
	\begin{align}
	\nonumber\varphi _{p,\ell ,s}^{\left( \text{LG} \right)}\left( r,\theta ,z \right)&=\sqrt{\frac{2p!}{\pi \left( p+\left| l \right| \right)!}}\cdot \frac{1}{w\left( z \right)}{{\left[ \frac{\sqrt{2}r}{w\left( z \right)} \right]}^{\left| \ell  \right|}}\exp \left[ -\frac{{{r}^{2}}}{{{w}^{2}}\left( z \right)} \right]L_{p}^{\left| \ell  \right|}\left[ \frac{2{{r}^{2}}}{{{w}^{2}}\left( z \right)} \right]\exp \left( \text{i}\ell \theta  \right) \\ \nonumber
	&\quad\times \exp \left[ \text{i}{{k}_{n,m,s}}z+\text{i}{{k}_{n,m,s}}\frac{{{r}^{2}}}{2R\left( z \right)}-\text{i}\left( 2p+\left| \ell  \right|+1 \right)\vartheta \left( z \right) \right] \\ \nonumber
	& \equiv \varphi _{p,\ell ,s}^{\left( \text{LG} \right)}\left( x,y,z \right) \\ \nonumber
	& =\sqrt{\frac{{{2}^{\left| \ell  \right|}}p!}{\pi \left( p+\left| l \right| \right)!}}\cdot \frac{1}{{{w}^{\left| \ell  \right|+1}}\left( z \right)}\exp \left[ -\frac{{{x}^{2}}+{{y}^{2}}}{{{w}^{2}}\left( z \right)} \right]L_{p}^{\left| \ell  \right|}\left[ 2\frac{{{x}^{2}}+{{y}^{2}}}{{{w}^{2}}\left( z \right)} \right]{{\left[ x+\text{i}\cdot \text{sgn}\left( \ell  \right)y \right]}^{\left| \ell  \right|}} \\ 
	& \quad\times \exp \left[ \text{i}{{k}_{n,m,s}}z+\text{i}{{k}_{n,m,s}}\frac{{{x}^{2}}+{{y}^{2}}}{2R\left( z \right)}-\text{i}\left( 2p+\left| \ell  \right|+1 \right)\vartheta \left( z \right) \right],
	\label{e6}  
	\end{align}
	where $L_{p}^{\ell}\left( \cdot  \right)$ represents the associated Laguerre polynomial with radial and azimuthal indices of $p$ and $\ell$. According to the formula for truncated diffraction of basic vortex, Eq.~(\ref{e2}), the truncated diffraction of nonplanar SU(2) mode as given by Eq.~(\ref{e5}) can be derived as:
	\begin{align}
	\nonumber{{E}_{\varphi }}\left( \xi ,\eta  \right)&=\int_{-\infty }^{\infty }{\int_{-\infty }^{\infty }{t\left( x,y \right)\Phi _{{{n}_{0}}}^{M}\left( \rho ,\theta ,z;{{\phi }_{0}}\left| \Omega  \right. \right)\text{exp}\left[ -\frac{\text{2}\pi \text{i}}{\lambda f}\left( \xi x+\eta y \right) \right]\text{d}x\text{d}y}} \\\nonumber 
	& =\frac{1}{{{2}^{{M}/{2}\;}}}\sum\limits_{K=0}^{M}{{{\left( \begin{matrix}
				M  \\
				K  \\
				\end{matrix} \right)}^{1/2}}}{{\text{e}}^{\text{i}K{{\phi }_{0}}}}\mathscr{F}\left[ t\left( x,y \right)\varphi _{0,{{n}_{0}}+Q\cdot K,{{s}_{0}}-P\cdot K}^{\left( \text{LG} \right)}\left( x,y,z \right) \right] \\ \nonumber
	& =\frac{1}{4{{\pi }^{2}}}\cdot \frac{ {{\text{e}}^{\text{i}{{k}_{{{n}_{0}},0,{{s}_{0}}}}z}}}{{{2}^{{M}/{2}\;}}}\cdot \sum\limits_{K=0}^{M}{\sqrt{\frac{{{2}^{{{n}_{0}}+QK}}p!}{\pi \left( p+{{n}_{0}}+QK \right)!}}\cdot \frac{1}{{{w}^{{{n}_{0}}+QK+1}}}{{\left( \begin{matrix}
				M  \\
				K  \\
				\end{matrix} \right)}^{1/2}}}{{\text{e}}^{\text{i}K{{\phi }_{0}}}}{{\text{e}}^{-\text{i}\left( {{n}_{0}}+QK+1 \right)\vartheta }} \\ 
	& \quad\times \mathscr{F}\left\{ t\left( x,y \right){{( x\pm \text{i}y )}^{{{n}_{0}}+QK}} \right\}*\mathscr{F}\left[ \exp \left( -\frac{{{x}^{2}}+{{y}^{2}}}{{{w}^{2}}} \right) \right]*\mathscr{F}\left[ L_{0}^{{{n}_{0}}+QK}\left( 2\frac{{{x}^{2}}+{{y}^{2}}}{{{w}^{2}}} \right) \right],
	\label{e7}
	\end{align}
	where we suppose that $R(z)$ is large enough to make approximation that $\exp \left[ \text{i}{{k}_{n,m,s}}\frac{{{x}^{2}}+{{y}^{2}}}{2R\left( z \right)} \right]\approx 1$, the term $\mathscr{F}\left\{ t\left( x,y \right){{( x\pm \text{i}y )}^{{{n}_{0}}+QK}} \right\}$ stands for the component of classical diffraction lattice, and the Fourier transform of the associated Laguerre polynomials should be Dirac delta functions because of $p=0$ here. Hereinafter, referring Eq.~(\ref{e2}) into Eq.~(\ref{e7}), we ignore the trivial amplitude-phase terms and continue the derivation:
	\begin{align}
	\nonumber{{E}_{\varphi }}\left( \xi ,\eta  \right)&\propto \sum\limits_{K=0}^{M}{\frac{{{2}^{{QK}/{2}\;}}{{\text{e}}^{\text{i}K{{\phi }_{0}}}}{{\text{e}}^{-\text{i}QK\vartheta }}}{{{w}^{QK}}\sqrt{\left( p+{{n}_{0}}+QK \right)!}}{{\left( \begin{matrix}
				M  \\
				K  \\
				\end{matrix} \right)}^{1/2}}}\mathscr{F}\left\{ t\left( x,y \right){{( x\pm \text{i}y )}^{{{n}_{0}}+QK}} \right\}*\mathscr{F}\left[ \exp \left( -\frac{{{x}^{2}}+{{y}^{2}}}{{{w}^{2}}} \right) \right] \\ \nonumber
	& \propto \sum\limits_{K=0}^{M}{\sum\limits_{k=0}^{{{n}_{0}}+QK}{\frac{{{2}^{{QK}/{2}\;}}{{\text{e}}^{\text{i}K{{\phi }_{0}}}}{{\text{e}}^{-\text{i}QK\vartheta }}}{{{w}^{QK}}\sqrt{\left( p+{{n}_{0}}+QK \right)!}} {{\left( \begin{matrix}
					M  \\
					K  \\
					\end{matrix} \right)}^{1/2}}\left( \begin{matrix}
			{{n}_{0}}+QK  \\
			k  \\
			\end{matrix} \right)}{{\left( \pm \text{i} \right)}^{k}}} \\ 
	& \quad\times \left[ \frac{{{\partial }^{{{n}_{0}}+QK}}}{\partial \omega _{x}^{{{n}_{0}}+QK-k}\partial \omega _{y}^{k}}T\left( {{\omega }_{x}},{{\omega }_{y}} \right) \right]*\exp \left( -{{w}^{2}}\frac{\omega _{x}^{2}+\omega _{y}^{2}}{4} \right),
	\label{e8}  
	\end{align}
	In this way, the theoretical diffraction lattices of the SU(2) vortices can be calculated associated with various given aperture function $t\left( x,y \right)$.
\end{widetext}
\section{Experimental design}
\begin{figure*}
	\centering
	\includegraphics[width=\linewidth]{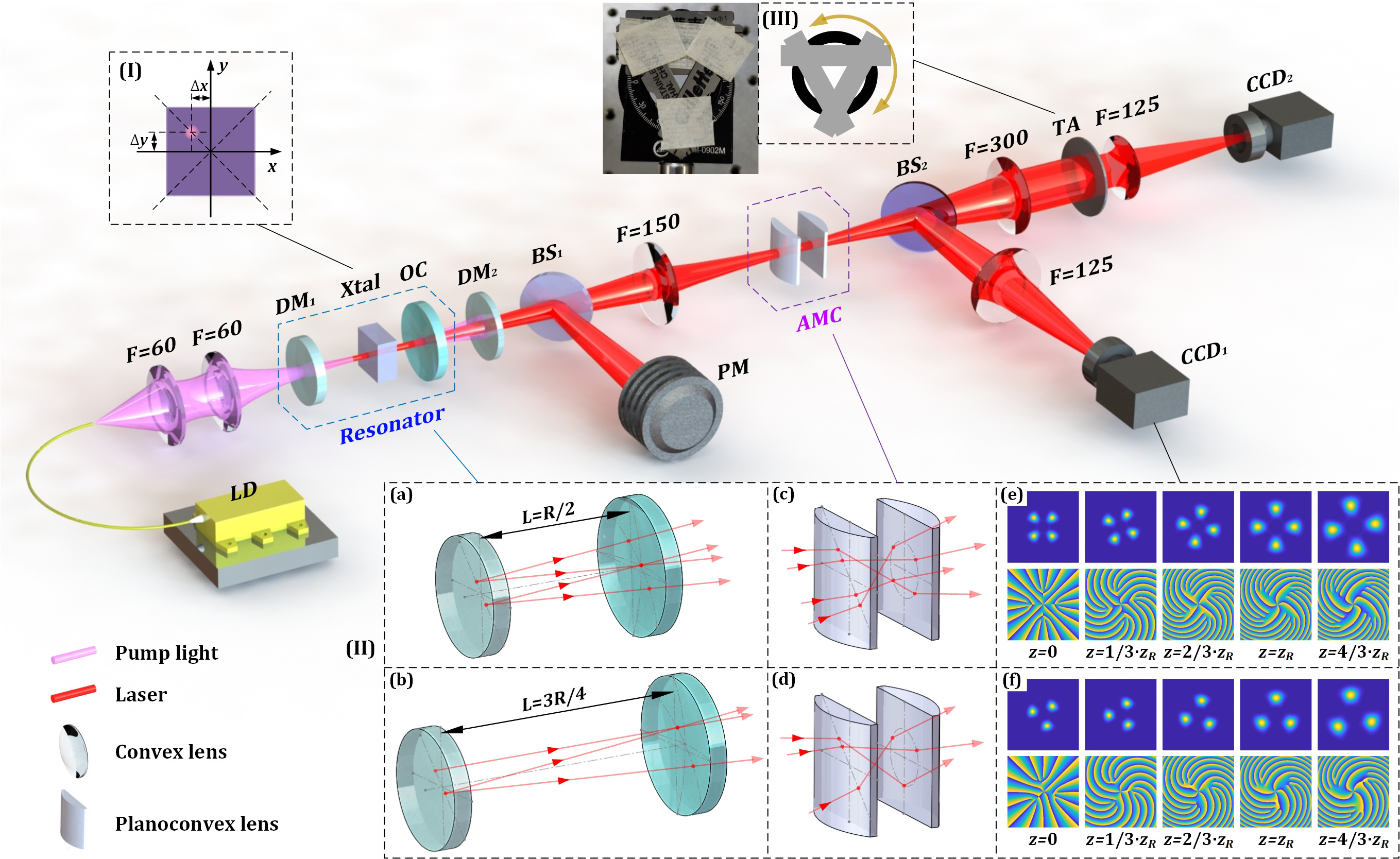}
	\caption{\label{f1} The experimental setup: LD, laser diode; DM, dichroic mirror; Xtal, crystal; OC, output coupler; BS, beam splitter; PM, power meter; AMC, astigmatic mode converter; TA, triangular aperture; CCD, charge coupled device. Inserts: (I) the controlling of pump spot on crystal; (II) the classical periodic ray trajectories in resonator under SU(2) degenerate states of (a) $|\Omega =1/4\rangle $ and (b) $|\Omega =1/3\rangle $, the classical ray trajectories in AMC for (c) $|\Omega =1/4\rangle $ and (d) $|\Omega =1/3\rangle $, and the theoretical intensity-phase profiles of SU(2) wave-packets versus $z$-axis for (e) $|\Omega =1/4\rangle $ and (f) $|\Omega =1/3\rangle $; (III) the details of the triangular aperture.}
\end{figure*}

The schematic of experimental setup is shown in Fig.~\ref{f1}. A diode-pumped solid-state laser was designed to generate planar SU(2) geometric mode. A high-power 976~nm fiber-coupled laser diode (LD) (Han's TCS, core: 105~$\upmu$m, NA: 0.22, highest power: 110~W) was used as the pump source. With two antireflective (AR) coated lenses with the focal length of $F=60$~mm, the pump light was focused into the laser crystal with the beam waist radius of about 200~$\upmu$m. The crystal is a slice-shape $4\times4\times2$~mm$^3$ Yb:CALGO (Altechna, a-cut, 5~at.\%, AR coated on end surfaces at 900--1100 nm), which was wrapped in a copper heat sink water-cooled at 18$^\circ$C. The pump coupling system was fixed at a precision adjusting stage to control the pump spot at 45$^\circ$-inclined off-axis position, as shown in inset (I) of Fig.~\ref{f1}, to generate 45$^\circ$-inclined planar geometric modes. The laser resonator was constituted by a plane dichroic mirror [DM$_1$, AR at 976 nm and high-reflective (HR) at 1030--1080~nm] and a concave output coupler (OC, transmittance of 1\% at 1020--1080~nm; $R = 100$~mm). Another dichroic mirror (DM$_2$, HR at 976~nm and AR at 1030--1080~nm) was used to filter the residual pump light. The cavity mirrors OC and DM$_1$ were installed on high-precision adjusting stages so that the cavity can be precisely adjusted to the frequency-degenerate states $|\Omega =1/4\rangle $ and $|\Omega =1/3\rangle $ at $L = R/2 = 50$~mm and $L = 3R/4 = 75$~mm, the classical periodic ray trajectories of which are respectively depicted in inset (II.a) and (II.b) of Fig.~\ref{f1}. 

Following the frontend, a beam splitter (BS$_1$, 45$^\circ$ incidence, T:R$\approx$1:99) was used to reflect the laser into a power meter to measure the power and the transmission light was used for mode conversion to obtain vortex SU(2) mode. Through a convex lens ($F=150$~mm), the laser was focused into the AMC constituted by two identical convex-plane cylindrical lenses with the focal length of 25~mm. The 45$^\circ$-inclined planar SU(2) mode was transformed into vortex SU(2) mode. The schematics of the transformation in AMC associated with the classical ray trajectories are illustrated in inset (II.c,d) of Fig.~\ref{f1}. Then, another beam splitter (BS$_2$, 45$^\circ$ incidence, T:R$\approx$1:1) was used to reflect the vortex beam into a camera (CCD$_1$) after being focused by a convex lens ($F = 125$~mm) for checking the quality of the generated vortex SU(2) wave-packet. The transmission light was collimated by a convex lens ($F = 300$~mm) and then illuminated on the homemade triangular aperture (TA). After being focused by a convex lens ($F = 125$~mm), the vortex beam was captured by another camera (CCD$_2$) for observing far-field diffraction pattern. The TA was installed on a rotating plate, the design and details of which are shown in inset (III) of Fig.~\ref{f1}. Thereby, the size, position, and orientation of the TA can all be controlled. We note that the convex lens with $F = 300$~mm could be axially displaced to adjust the range and orientation of the SU(2) mode relative to the TA, because the SU(2) wave-packet has the range-varying and twisting effects along the $z$-axis propagation led by the macroscopic performance of OAM. The corresponding theoretical intensity and phase profiles for $|\Omega =1/4\rangle$ ($n_0=4$, $M=0$, $\phi_0=0$) and $|\Omega =1/3\rangle$ ($n_0=4$, $M=0$, $\phi_0=1.2\pi$) states are depicted in insets (II.e) and (II.f) of Fig.~\ref{f1}. Therefore, this above presented system can be applied to flexibly investigate the truncated diffraction patterns of vortex SU(2) geometric modes with various conditions of TA. 

As can be seen in the theoretical vortex SU(2) wave-packets, in contrast to classical doughnut-shape LG mode, there are multiple splitting singularities that exist not only in the center, giving rise to a dark region, but also in the partial dark regions. It has been illustrated in literature that, for the SU(2) wave-packet $\Phi _{{{n}_{0}}}^{M}$, the topological charges of the center singularity are related by $n_0$, and the topological charges in each partial dark regions are related by $M$. It is our target to investigate the peculiar properties of the truncated diffraction lattices related to the special singularities distributions under specially arranged triangle apertures.

\begin{figure*}[htbp]
	\centering
	\includegraphics[width=0.98\linewidth]{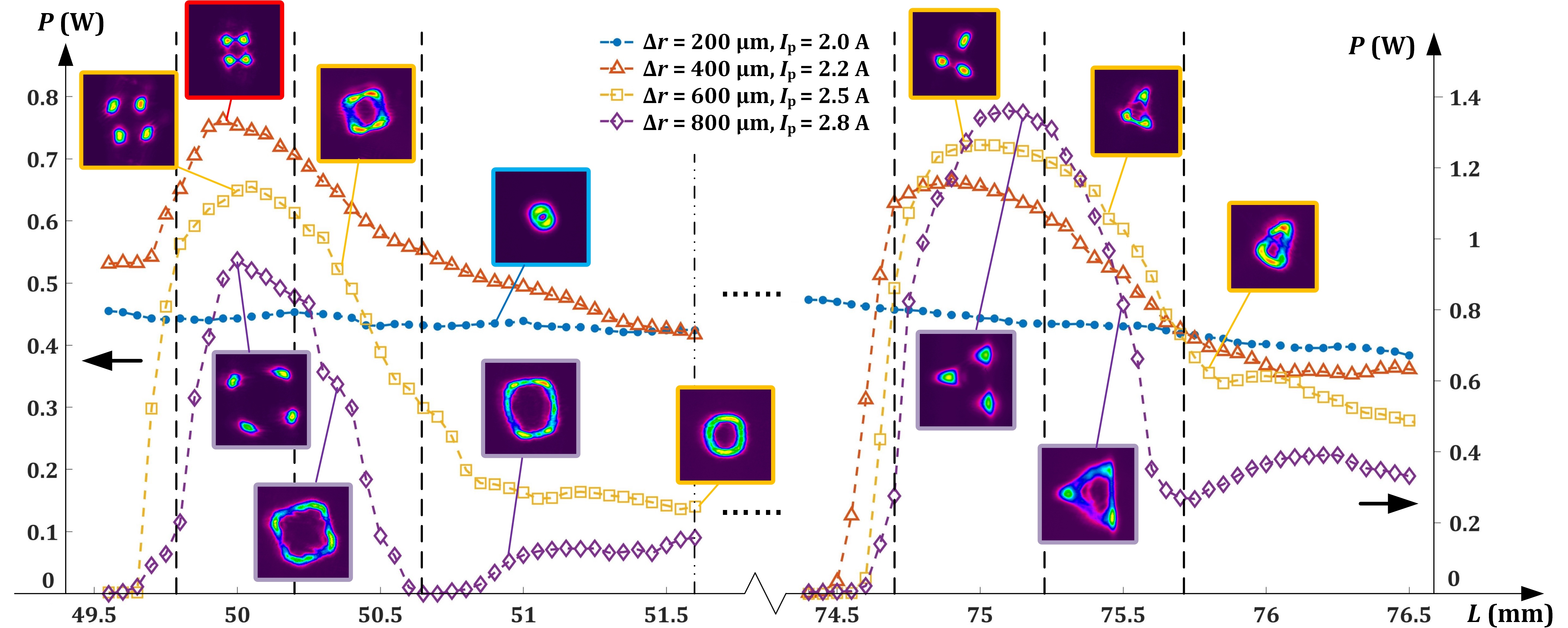}
	\caption{\label{f2} The experimental output power curves under different pump currents off-axis displacements, associated with the corresponding transverse patterns.}
\end{figure*}
\begin{figure*}[htbp]
	\centering
	\includegraphics[width=0.55\linewidth]{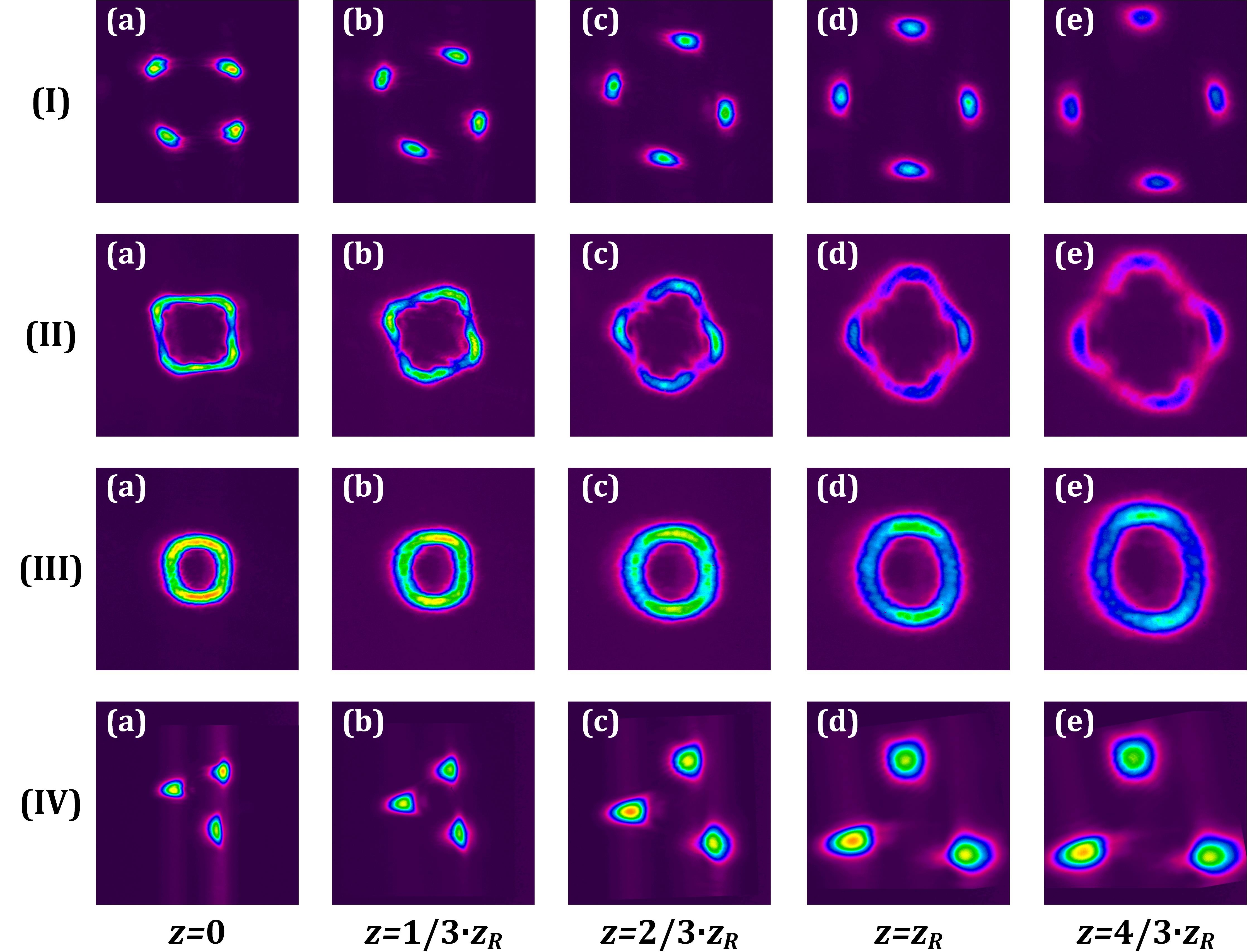}
	\caption{\label{f3} The experimental nonplanar SU(2) mode patterns evolution with different cavity length from $|\Omega =1/\text{4}\rangle $ to $|\Omega =1/\text{3}\rangle $ [from (I)-(IV)] at various propagation distances (a) $z = z_R$, (b) $z = 1/3\cdot z_R$, (c) $z = 2/3\cdot z_R$, (d) $z = z_R$, (e) $z = 4/3\cdot z_R$.}
\end{figure*}
\section{Results and discussions}
\subsection{Generation and control of SU(2) wave-packet}
Due to the characteristics of ray-wave duality, the SU(2) mode is always generated with a drastic increase of output power at corresponding degenerate states in cavity \cite{16,17}. In our experiment, we found two observable power peaks versus cavity length at states $|\Omega =1/4\rangle $ ($L = R/2 = 50$~mm) and $|\Omega =1/3\rangle $ ($L = 3R/4 = 75$~mm), under certain off-axis degrees and pump powers in our Yb:CALGO laser. 

As is seen in Fig.~\ref{f2}, under off-axis displacement ($\Delta r=\sqrt{\Delta {{x}^{2}}+\Delta {{y}^{2}}}$) of 200~$\upmu$m and pump current of $I_{\rm p} = 2.0$~A (pump power $P_{\rm p}=16.8$~W), the laser mode (captured by CCD$_1$) always performs as a low-order doughnut-shape vortex beam; when $\Delta r = 400$~$\upmu$m and $I_{\rm p} = 2.2$~A ($P_{\rm p}=18.6$~W), the power peaks occur at the corresponding degenerate states and the clearest SU(2) wave-packets can be captured; when $\Delta r = 600$~$\upmu$m and $I_{\rm p} = 2.5$~A ($P_{\rm p}=22.8$~W), the power peaks effects maintain while the ranges of the SU(2) wave-packets become larger, i.e. the mode orders $n_0$ and $M$ are higher; when $\Delta r = 800$~$\upmu$m and $I_{\rm p} = 2.8$~A ($P_{\rm p}=27.3$~W), the orders of the SU(2) wave-packets become even larger. Generally, with the higher degree of pump off-axis deviation, the range of the generated SU(2) wave-packet is more enlarged, corresponding to the higher mode orders. 

Additionally, when the cavity length was gradually deviating from a degenerate state, the SU(2) wave-packet gradually transformed into polygonal vortex beam \cite{25} and then into doughnut-shape LG mode. As another important property of nonplanar SU(2) mode, the wave-packet can twist along the propagation distance. The twisting effect were also checked in our vortex beams, as shown in Fig.~\ref{f3}, which reveals the macroscopic existence of OAM of light. Summarily, we can generate various SU(2) wave-packet $\Phi _{{{n}_{0}}}^{M}$ and classical LG vortices with different orders via pump control in our system. 

\begin{figure*}
	\centering
	\includegraphics[width=0.88\linewidth]{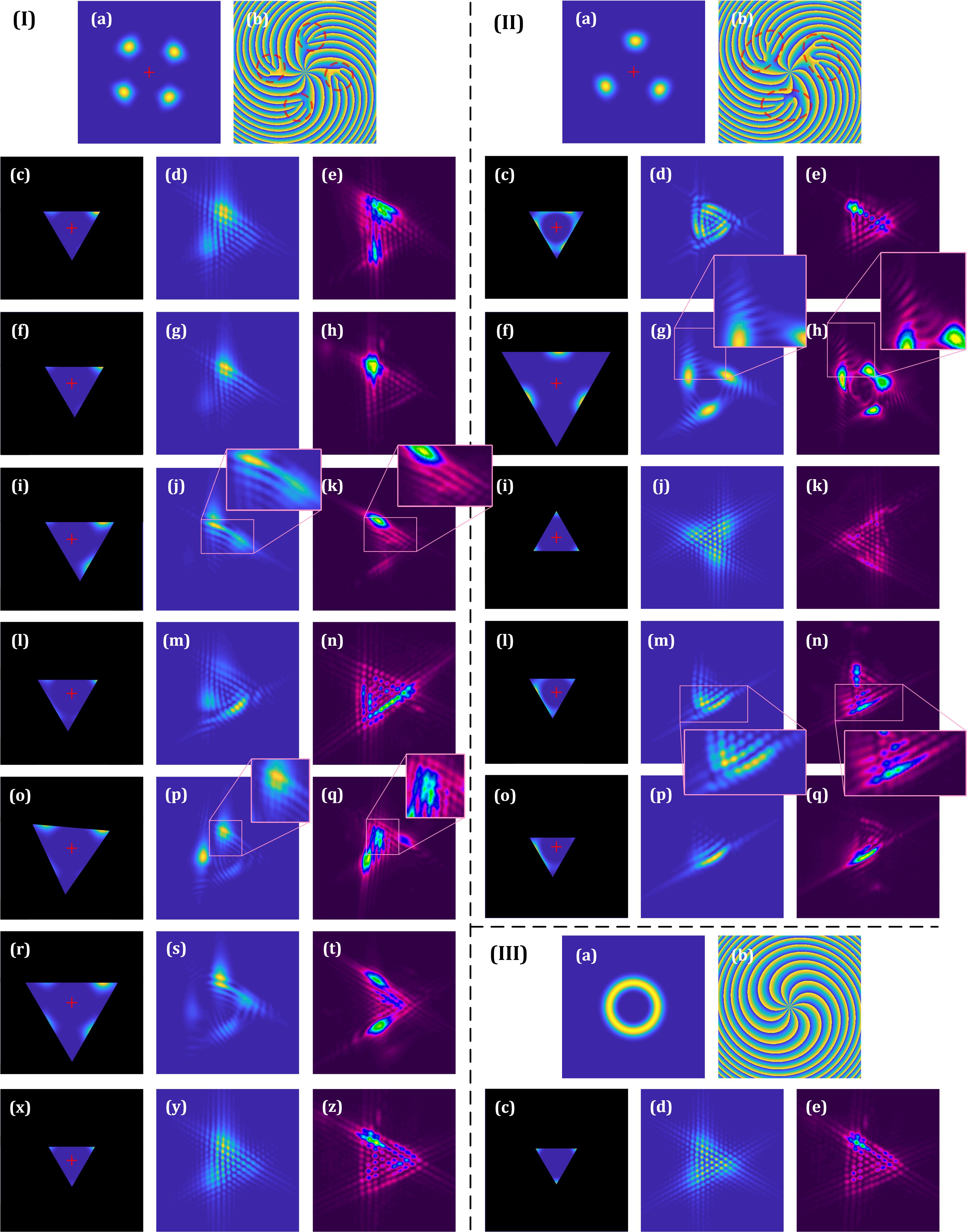}
	\caption{\label{f4} The theoretical and experimental results of far-field truncated diffraction for (I) a $|\Omega =1/\text{4}\rangle $ SU(2) mode, (II) a $|\Omega =1/\text{3}\rangle $ SU(2) mode and (III) a LG mode [(a): intensity; (b): phase], including the conditions of TA (left columns), simulated patterns (middle columns), and experimental patterns (right columns). The symbols of + mark the center origin of coordinates and the dashed circles mark the partial OAM regions.}
\end{figure*}
\section{Unveiling meshy and spot-array optical lattices}
Figure~\ref{f4} shows the theoretical and experimental results of far-field truncated diffraction for (I) a $\Phi _{{{n}_{0}}}^{M}{{\big|}_{\Omega ={1}/{4}\;}}$ mode, (II) a $\Phi _{{{n}_{0}}}^{M}{{\big|}_{\Omega ={1}/{3}\;}}$ mode, and (III) an LG mode, associated with various conditions of TA (left columns: the conditions of TA; middle columns: the simulated patterns; right columns: the experimental patterns). 

For the $\Phi _{{{n}_{0}}}^{M}{{\big|}_{\Omega ={1}/{4}\;}}$ mode [Fig.~\ref{f4}(I.a, I.b)], we firstly installed a TA with similar size as that of center dark region [Fig.~\ref{f4}(I.c)]. An optical lattice pattern was observed, yet with a very nonuniform intensity distribution in contrast to the classical spot-array lattices. The intensity concentrated region performed more like a muddy meshy structure rather than the spot-array structure. Despite that the centers of the TA and $\Phi _{{{n}_{0}}}^{M}{{\big|}_{\Omega ={1}/{4}\;}}$ wave-packet were kept the same, this asymmetry is still understandable, because the intrinsic symmetries of TA and $\Phi _{{{n}_{0}}}^{M}{{\big|}_{\Omega ={1}/{4}\;}}$ are different, i.e. the symmetry of TA belongs to D$_6$ (dihedral group) while $\Phi_{{{n}_{0}}}^{M}{{\big|}_{\Omega ={1}/{4}\;}}$ belongs to D$_8$. When we slightly moved the position of the TA, the condition of intensity concentration correspondingly changed [Fig.~\ref{f4}(I.c-l)].  Particularly, when the TA covered more partial OAM region, the meshy structure could be more salient in one direction [Fig.~\ref{f4}(I.i-k)]. Furthermore, when we enlarged the size and rotate the orientation of TA to cover more partial OAM region, the concentrated meshy structures were easier to emerge [Fig.~\ref{f4}(I.o-t)]. On the contrary, when we shrinked the size of TA to nearly only cover the center OAM region, the diffraction pattern [Fig.~\ref{f4}(I.x-z)] was much close to the classical truncated triangular lattice [Fig.~\ref{f4}(III)]. 

We did the same experiments for the $\Phi _{{{n}_{0}}}^{M}{{\big|}_{\Omega ={1}/{3}\;}}$ wave-packet. When the TA had a similar size as the center dark region, the diffraction lattices pattern always showed some periodic meshy structures [Fig.~\ref{f4}(II.c-e)]. When the TA was enlarged to cover more partial OAM region, the salient meshy structures would emerge [Fig.~\ref{f4}(II.f-h)]. In this case, note that the TA and $\Phi _{{{n}_{0}}}^{M}{{\big|}_{\Omega ={1}/{3}\;}}$ have the same symmetry of D$_6$, thus, the three partial OAM regions can all be covered and the meshy structures emerge correspondingly at three parts [Fig.~\ref{f4}(II.i)]. Moreover, the size and orientation of TA also play important roles in approximating the classical lattices. To this end, the size should be small enough and the orientation should be rotated to a proper angle for avoiding the partial dark region [Fig.~\ref{f4}(II.i-k)]. Figures.~\ref{f4}(II.l) and \ref{f4}(II.o) are under the same TA size as that of Fig.~\ref{f4}(II.i) but in different angle and position. Their corresponding diffraction lattices also comprise concentrated meshy structures [Fig.~\ref{f4}(II.l-q)]. According to the above results, the meshy structures in diffraction lattices should be closely correlated by the special partial OAM region of SU(2) wave-packet. Since the partial OAM is also interrelated with the mode indices $n_0$ and $M$, the heuristic arguments inspire us to exploit a powerful application: a direct way to measure the overall OAM information, the topological charges with singularities distributions, of a vortex SU(2) wave-packet.

\begin{figure*}
	\centering
	\includegraphics[width=0.9\linewidth]{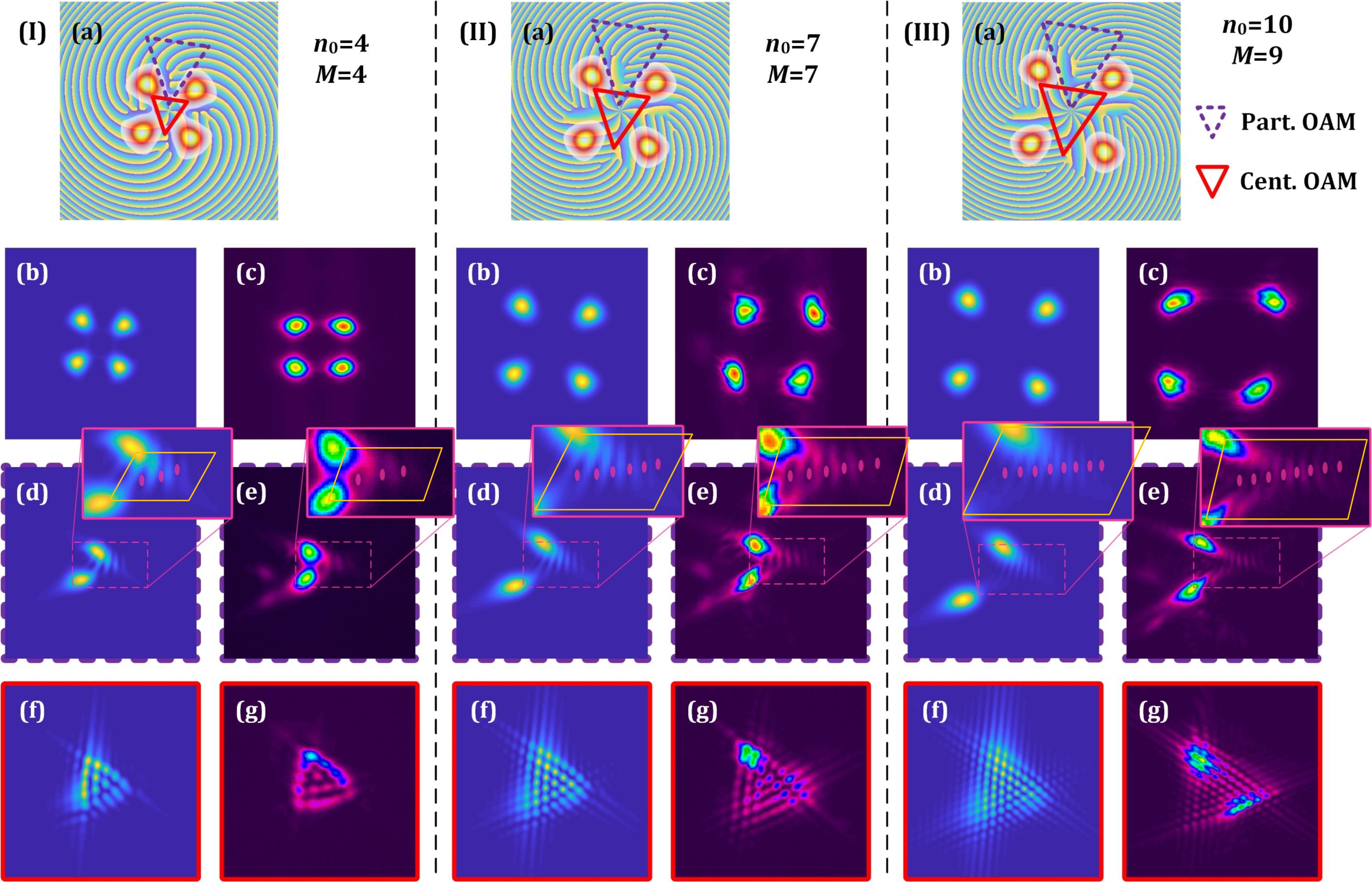}
	\caption{\label{f5} The OAM detection of SU(2) wave-packets (I) $\Phi _{4}^{4}{{\big|}_{\Omega ={1}/{4}\;}}$, (II) $\Phi _{7}^{7}{{\big|}_{\Omega ={1}/{4}\;}}$, and (III) $\Phi _{10}^{9}{{\big|}_{\Omega ={1}/{4}\;}}$, including (a) the schematics of center and partial OAM detection, (b,c)  the simulated and experimental intensity field, (d,e) the simulated and experimental truncated diffraction lattices for partial OAM detection, (f,g) the simulated and experimental truncated diffraction lattices for center OAM detection.}
\end{figure*}
\section{Detecting the center and partial OAM of an SU(2) wave-packet}
For an integer-order OAM vortex, the singularity with an integer topological charge at the center gives rise to a dark hole, as a result the average OAM per photon is $m\hbar$ where $m$ is an integer. For a fractional OAM vortex, the average OAM per photon is $q\hbar$ where $q$ is a fraction due to the singularities split into other region outside the center singularity \cite{26a,26b}. It has been demonstrated that SU(2) vortices are fractional OAM vortices, because there are not only a center singularity but also singularities existed in partial region \cite{19}. Therefore, it can reveal the OAM information via measuring the topological charges of the singularity at the center and that of the singularities in the partial region respectively.

Theoretically, except the center singularity with charge of $n_0$ for $\Phi _{{{n}_{0}}}^{M}{{\big|}_{\Omega ={P}/{Q}\;}}$, the multiple singularities in a field are always split into $Q$ parts of partial dark region with charge of $M$. The whole field shows singularities with number of $n_0 + QM$ leading to a large fractional OAM. However, the classical interferometry cannot identify the actual phase and topological charges of SU(2) vortices, because the exotic intensity fields forbid the overall emergence of interference fringes \cite{18}. We hereinafter introduce an effective method to detect the OAM in both center and partial regions for SU(2) vortices. 

As shown in Fig.~\ref{f5}(I), we firstly generated an SU(2) wave-packet analogous to the theoretical prediction of $\Phi _{4}^{4}{{\big|}_{\Omega ={1}/{4}\;}}$ [theory: Fig.~\ref{f5}(I.a,b); experiment: Fig.~\ref{f5}(I.c)], and then detected its truncated diffraction pattern. When we adjusted the TA to cover a partial OAM region [like the dashed line noted in Fig.~\ref{f5}(I.a)], the meshy structure could be more and more concentrated and finally emerge as a dark-hole array structure. Then, we can count the holes to evaluate the topological charge at the partial region. Additionally, the holes intensity distributions should both be considered during the counting process. As can be seen, there are two bright spots and the clear holes with number of 3 can be counted at the right side of the two bright spots [Fig.~\ref{f5}(I.d,e)]. When we adjusted the TA to the center avoiding the partial OAM region, the center charge could be measured by the classical principle \cite{10,11}, i.e. the side row of the optical lattice carries clear bright spots with number of 5. In the case of $\Phi _{7}^{7}{{\big|}_{\Omega ={1}/{4}\;}}$ [Fig.~\ref{f5}(II.a-c)], the meshy structure shows clear dark holes with number of 6 at the right side of two bright spots [Fig.~\ref{f5}(II.d-e)] and the classical lattice pattern shows clear bright spots at side row with number of 8. The experiment was also done for an even high-order case $\Phi _{10}^{9}{{\big|}_{\Omega ={1}/{4}\;}}$ [Fig.~\ref{f5}(III.a-c)], the meshy structure shows clear dark holes with number of 8 at the right side of two bright spots [Fig.~\ref{f5}(III.d-e)] and the classical lattice shows clear bright spots at side row with number of 11. Summarily, for an SU(2) wave-packet $\Phi _{{{n}_{0}}}^{M}{{\big|}_{\Omega ={P}/{Q}\;}}$ under a proper control of TA, the meshy structure shows clear dark holes with number of $M-1$, which reveals a topological charge of $M$ in each partial dark region; the classical lattice shows clear bright spots at side row with number of ${{n}_{0}}+1$, which reveals a topological charge of $n_0$ in each partial dark region.  Finally, as an expectation, we note that this method can be further referred to measure more kinds of others complex SU(2) wave-packets such as Trochoidal wave-packets \cite{26,27}, Lissajous wave-packets \cite{28}, hyperboloid polarized wave-packets \cite{29,30}, multi-axis vortices \cite{31,32}, and symmetry-breaking SU(2) wave-packets \cite{32}. We note that it is tolerable that there might be $\pm 1$ error in holes counting process due to the experimental error. It is common that there is actual error of device alignment, apture control and so on between the experimental and theoretical results, which however does not affect the quantitative evaluation method in actual applications.

\section{Summary}
We for the first time, to the best of our knowledge, demonstrate the truncated diffraction lattices with TA of the SU(2) geometric modes, the principle of which is elaborately utilized to detect the OAM of vortex SU(2) wave-packet $\Phi _{{{n}_{0}}}^{M}{\big|_{\Omega ={P}/{Q}}}$. Through theoretical and experimental studies of far-field truncated diffraction, a meshy structure is unveiled by adjusting the illuminated TA, which is correlated by the singularities splitting of $\Phi _{{{n}_{0}}}^{M}{{\big|}_{\Omega ={P}/{Q}}}$. For measuring the topological charge in the partial OAM regions, the TA should be arranged to cover corresponding partial dark region, and the meshy structure in diffraction pattern shows clear dark holes with number of $M-1$. For the center OAM, the TA should cover the center and avoid the partial dark region, and then the diffraction lattice can show clear bright spots at side row with number of ${{n}_{0}}+1$. These effects are fully validated by theoretical simulations, which greatly extend the versatility of measuring the topological structures of special beams.

\end{document}